\documentclass[a4paper, final, natbib]{svjour3}    

\usepackage[english]{babel}
\usepackage{hyperref}
\usepackage{graphicx}
\usepackage{amsmath, amsbsy, amsfonts}
\usepackage{multirow}
\usepackage{lscape}
\usepackage{rotating}

\def\Rbb{\mathbb{R}}

\def\Zbb{\mathbb{Z}}

\def\Ascr{\mathcal{A}}
\def\Bscr{\mathcal{B}}
\def\Cscr{\mathcal{C}}
\def\Dscr{\mathcal{D}}
\def\Escr{\mathcal{E}}

\def\Gscr{\mathcal{G}}
\def\Hscr{\mathcal{H}}
\def\Iscr{\mathcal{I}}

\def\Lscr{\mathcal{L}}

\def\Oscr{\mathcal{O}}

\def\Rscr{\mathcal{R}}
\def\Sscr{\mathcal{S}}
\def\Tscr{\mathcal{T}}

\def\avec{\mathbf{a}}

\def\evec{\mathbf{e}}

\def\kvec{\mathbf{k}}
\def\lvec{\mathbf{l}}
\def\mvec{\mathbf{m}}
\def\nvec{\mathbf{n}}

\def\rvec{\mathbf{r}}

\def\xvec{\mathbf{x}}
\def\yvec{\mathbf{y}}
\def\Lambdavec{\boldsymbol{\Lambda}}
\def\lambdavec{\boldsymbol{\lambda}}
\def\xivec{\boldsymbol{\xi}}
\def\etavec{\boldsymbol{\eta}}
\def\nuvec{\boldsymbol{\nu}}
\def\Lvec{\mathbf{L}}
\def\Ivec{\mathbf{I}}

\def\phivec{\boldsymbol{\phi}}
\def\rhovec{\boldsymbol{\rho}}
\def\omegavec{\boldsymbol{\omega}}
\def\dispchi{\raise2pt\hbox{$\chi$}}
\def\scriptchi{\raise1pt\hbox{$\scriptstyle\chi$}}
\def\rtvec{\tilde\rvec}
\def\lie#1{\Lscr_{#1}}
\def\epsilon{\varepsilon}
\def\rho{\varrho}
\def\phi{\varphi}
\def\csi{\xi}
\def\build#1_#2^#3{\mathrel{
\mathop{\kern 0pt#1}\limits_{#2}^{#3}}}
\newlength{\arrayrulewidthOriginal}
\newcommand{\Cline}[2]{%
  \noalign{\global\setlength{\arrayrulewidthOriginal}{\arrayrulewidth}}%
  \noalign{\global\setlength{\arrayrulewidth}{#1}}\cline{#2}%
  \noalign{\global\setlength{\arrayrulewidth}{\arrayrulewidthOriginal}}}

\date{Received: 9 November 2012 / Accepted: 12 June 2013}
\begin{document}

\title{On the extension of the Laplace-Lagrange secular theory to
  order two in the masses for extrasolar systems}

\author{Anne-Sophie Libert \and Marco Sansottera}

\institute{A.-S. Libert \at
  naXys, Department of Mathematics, University of Namur,
  8 Rempart de la Vierge,\\
  5000~Namur, Belgium \\
  Observatoire de Lille (LAL-IMCCE), CNRS-UMR8028,
  1 Impasse de l'Observatoire,\\
  59000~Lille, France\\
  \email{anne-sophie.libert@fundp.ac.be}\\ 
  \and M. Sansottera \at
  naXys, Department of Mathematics, University of Namur,
  8 Rempart de la Vierge,\\
  5000~Namur, Belgium \\
  \email{marco.sansottera@fundp.ac.be}}        
\maketitle

\begin{abstract}
We study the secular evolution of several exoplanetary systems by
extending the Laplace-Lagrange theory to order two in the masses.
Using an expansion of the Hamiltonian in the Poincar\'e canonical
variables, we determine the fundamental frequencies of the motion and
compute analytically the long-term evolution of the keplerian
elements.  Our study clearly shows that, for systems close to a
mean-motion resonance, the second order approximation describes their
secular evolution more accurately than the usually adopted first order
one.  Moreover, this approach takes into account the influence of the
mean anomalies on the secular dynamics.  Finally, we set up a simple
criterion that is useful to discriminate between three different
categories of planetary systems: (i)~{\it secular} systems (HD~11964,
HD~74156, HD~134987, HD~163607, HD~12661 and HD~147018); (ii)~systems
{\it near a mean-motion resonance} (HD~11506, HD~177830, HD~9446,
HD~169830 and $\upsilon$~Andromedae); (iii)~systems {\it really close
  to} or {\it in a mean-motion resonance} (HD~108874, HD~128311 and
HD~183263).

\end{abstract}

\section{Introduction}\label{sec:intro}
The study of the dynamics of planetary systems is a long standing and
challenging problem. The classical perturbation theory, mainly
developed by Lagrange and Laplace, uses the circular approximation as
a reference for the orbits. The discovery of extrasolar planetary
systems has opened a new field in Celestial Mechanics and nowadays
more than $100$ multi-planetary systems have been discovered. In
contrast to the Solar system, where the orbits of the planets are
almost circular, the exoplanets usually describe true ellipses with
high eccentricities. Thus, the applicability of the classical
approach, using the circular approximation as a reference, can be
doubtful for these systems. In this work we revisit the classical
Laplace-Lagrange theory for the secular motions of the pericenters of
the planetary orbits, based only on a linear approximation of the
dynamical equations, by considering also higher order terms.

Previous works of \cite{LibHen-2005,LibHen-2006} for coplanar systems
have generalized the classical expansion of the perturbative potential
to a higher order in the eccentricities, showing that this analytical
model gives an accurate description of the behavior of planetary
systems which are not close to a mean-motion resonance, up to
surprisingly high eccentricities. Moreover, they have shown that an
expansion up to order $12$ in the eccentricities is usually required for
reproducing the secular behavior of extrasolar planetary systems.
This expansion has also been used by \cite{BeaNesDon-2006} to
successfully reproduce the motions of irregular satellites with
eccentricities up to~$0.7$. \cite{VerArm-2007} have highlighted the
limitations of lower order expansions; using only a fourth-order
expansion in the eccentricities, they did not reproduce, even
qualitatively, the secular dynamics of extrasolar planetary
systems. All the previous results have been obtained considering a
secular Hamiltonian at order one in the masses. Let us also remark
that an alternative {\it octupole-level} secular theory has been
developed for systems that exhibit {\it hierarchical behavior} (see,
e.g.,~\cite{ForKozRas-2000}, \cite{LeePea-2003},
\cite{NaoFarLitRasTey-2011}, \cite{KatDonMal-2011} and
\cite{LibDel-2012}). However, this approach is not suitable for our
study, as we will also consider systems with large semi-major axes
ratio.

Considering the secular dynamics of the Solar system, Lagrange and
Laplace showed that the proximity to the $5$:$2$ mean-motion resonance
between Jupiter and Saturn leads to large perturbations in the secular
motion, explaining the so-called ``great inequality''.  Let us stress
that when referring to {\it secular evolution} we mean long-term
evolution, possibly including the long-term effects of near
resonances.  Indeed the terms of the perturbation associated to
mean-motion resonances have small frequencies and thus influence the
secular behavior of the system.  A good description of the secular
dynamics of an exoplanetary system should include the effects of the
mean-motion resonances on their secular long-term
evolution. Therefore, we replace the classical circular approximation
with a torus which is invariant up to order two in the masses, this is
the so-called Hamiltonian at order two in the masses.  The benefit of
a second order approach has been clearly highlighted in
\cite{Laskar-1988}, see Table 2 in that paper, where a comparison
between the fundamental frequencies of the planetary motion of the
Solar system, is given for different approximations.  Concerning the
problem of the stability of the Solar system, the celebrated theorems
of Kolmogorov and Nekhoroshev allowed to make substantial progress.
Nevertheless, in order to apply these theorems, a crucial point is to
consider the secular Hamiltonian at order two in the masses in order
to have a good approximation of the secular dynamics (e.g.,~this
allows, in \cite{LocGio-2007}, to deal with the true values of the
planetary masses, while the first order approximation used in
\cite{LocGio-2005} forces the authors to reduce the masses of the
planets by a factor 10).  In recent years, the estimates for the
applicability of both Kolmogorov and Nekhoroshev theorems to realistic
models of some part of the Solar system have been improved by some
authors (e.g.,~\cite{Robutel-1995}, \cite{Fejoz-2005},
\cite{CelChi-2005}, \cite{GabJorLoc-2005}, \cite{LocGio-2007},
\cite{GioLocSan-2009} and \cite{SanLocGio-2011a, SanLocGio-2011b}).

In the present contribution, we study the secular dynamics of
extrasolar planetary systems consisting of two coplanar planets. The
aim is to reconstruct the evolution of the eccentricities and
pericenters of the planets by using analytical techniques, extending
the Laplace-Lagrange theory to order two in the masses. To do so, we
extend the results in \cite{LibHen-2005,LibHen-2006}, replacing the
first order averaged Hamiltonian, with the one at order two in the
masses, and show the improvements of this approximation on the study
of the secular evolution of extrasolar systems. In particular, we
determine the fundamental frequencies of the motion and compute
analytically the long-term evolution of the keplerian
elements. Furthermore, we show that the Hamiltonian at order two in
the masses describes accurately the secular dynamics of systems close
to a mean-motion resonance and, as a byproduct, we also give an
estimate of the proximity to a mean-motion resonance of the two-planet
extrasolar systems discovered so far.

The paper is organized as follows. In Section~\ref{sec:expansion}, we
describe the expansion of the Hamiltonian of the planar three-body
problem in Poincar\'e variables.  Following the Lagrange approach, we
focus, in Section~\ref{sec:secular}, on the secular part of the
Hamiltonian and derive the secular Hamiltonian at order two in the
masses. In Section~\ref{sec:actang}, we construct a high-order
Birkhoff normal form, using the Lie series method, that leads to a
very simple form of the equations of motion, being function of the
actions only. We also show how to compute the secular frequencies and
perform a long-term analytical integration of the motion of the
planets. In Section~\ref{sec:andromedae}, we apply our method to the
$\upsilon$ Andromedae system and show that the second order
approximation is well suited for systems close to a mean-motion
resonance. Furthermore, the influence of the mean anomaly on the
long-term evolution is pointed out in Section~\ref{sec:M}. In
Section~\ref{sec:proxy}, we set up a criterion to evaluate the
proximity of planetary systems to mean-motion resonances, and apply it
to the two-planet extrasolar systems discovered so far. Finally, our
results are summarized in Section~\ref{sec:results}.  An appendix
containing the expansion of the secular Hamiltonian of the $\upsilon$
Andromedae extrasolar system, up to order $6$, follows.

\section{Expansion of the planetary Hamiltonian}\label{sec:expansion}
We consider a system of three coplanar point bodies, mutually
interacting according to Newton's gravitational law, consisting of a
central star $P_0$ of mass $m_0$ and two planets $P_1$ and $P_2$ of
mass $m_1$ and $m_2$, respectively. The indices $1$ and $2$ refer to
the inner and outer planet, respectively.

Let us now recall how the classical Poincar\'e variables can be
introduced to perform the expansion of the Hamiltonian around circular
orbits. We follow the formalism introduced by Poincar\'e (see
\cite{Poincare-1892, Poincare-1905}; for a modern exposition, see,
e.g.,~\cite{Laskar-1989a} and \cite{LasRob-1995}). To remove the
motion of the center of mass, we adopt the heliocentric
coordinates\footnote{Let us note that the Jacobi variables are less
  suitable for our purpose, as they require a Taylor expansion in the
  planetary masses.}, $\rvec_j=\overrightarrow{P_0P_j}$, with $j =
1,\, 2$.  Denoting by $\rtvec_j$ the momenta conjugated to $\rvec_j$,
the Hamiltonian of the system has four degrees of freedom, and reads
\begin{equation}
F(\rvec,\rtvec)=
T^{(0)}(\rtvec)+U^{(0)}(\rvec)+
T^{(1)}(\rtvec)+U^{(1)}(\rvec)\ ,
\label{eq:H_iniz}
\end{equation}
where
\begin{alignat*}{2}
T^{(0)}(\tilde{\bf r})&=\frac{1}{2}\sum_{j=1}^2 \|\tilde{\bf r}_j\|^2 \left(\frac{1}{m_0}+\frac{1}{m_j} \right)\ , \qquad&
T^{(1)}(\tilde{\bf r})&= \frac{\tilde{\bf r}_1 \cdot \tilde{\bf r}_2}{m_0}\ ,\\
U^{(0)}({\bf r})&=-\Gscr\sum_{j=1}^2 \frac{m_0m_j}{\|{\bf r}_j\|}\ , \qquad&
U^{(1)}({\bf r})&= -\Gscr\frac{m_1m_2}{\|{\bf r}_1-{\bf r}_2\|}\ .
\end{alignat*}

The plane set of Poincar\'e canonical variables is introduced as
\begin{equation}
\vcenter{\openup1\jot\halign{
 \hbox {\hfil $\displaystyle {#}$}
&\hbox {\hfil $\displaystyle {#}$\hfil}
&\hbox {$\displaystyle {#}$\hfil}
&\hbox to 6 ex{\hfil$\displaystyle {#}$\hfil}
&\hbox {\hfil $\displaystyle {#}$}
&\hbox {\hfil $\displaystyle {#}$\hfil}
&\hbox {$\displaystyle {#}$\hfil}\cr
\Lambda_j &=& \frac{m_0\, m_j}{m_0+m_j}\sqrt{\vphantom{b^a}\Gscr(m_0+m_j) a_j}\ ,
& &\lambda_j &=& M_j+\omega_j\ ,
\cr
\csi_j &=& \sqrt{\vphantom{b^a} 2\Lambda_j}\,
\sqrt{1-\sqrt{1-e_j^2}}\,\cos\omega_j\ ,
& &\eta_j&=&-\sqrt{\vphantom{b^a}2\Lambda_j}\,
\sqrt{1-\sqrt{1-e_j^2}}\, \sin\omega_j\ ,
\cr
}}
\label{eq:poincvar}
\end{equation}
for $j=1\,,\,2\,$, where $a_j\,,\> e_j\,,\> M_j$ and $\omega_j$ are
the semi-major axis, the eccentricity, the mean anomaly and the
longitude of the pericenter of the $j$-th planet, respectively.  One
immediately sees that both $\xi_j$ and $\eta_j$ are of the same order
as the eccentricity $e_j$.  Using the Poincar\'e canonical variables,
the Hamiltonian becomes
\begin{equation}
F(\Lambdavec, \lambdavec, \xivec, \etavec)=
F^{(0)}(\Lambdavec)+F^{(1)}(\Lambdavec, \lambdavec, \xivec
,\etavec)\ ,
\label{eq:H_iniz_poinc}
\end{equation}
where $F^{(0)}=T^{(0)}+U^{(0)}$ is the keplerian part and
$F^{(1)}=T^{(1)}+U^{(1)}$ the perturbation. Let us emphasize that the
ratio $F^{(1)}/F^{(0)}=\Oscr(\mu)$ with
$\mu=\max\{m_1/m_0,m_2/m_0\}$. Therefore, the time derivative of each
variable is of order $\mu$, except for $\lambdavec$. For this reason
we will refer to $(\Lambdavec,\lambdavec)$ as the {\it fast variables} and
to $(\xivec,\etavec)$ as the {\it secular variables}.

We proceed now by expanding the Hamiltonian~\eqref{eq:H_iniz_poinc} in
Taylor-Fourier series.  We pick a fixed value $\Lambdavec^*$ of the
fast actions\footnote{We recall that, as shown by Poisson, the
  semi-major axes are constant up to the second order in the
  masses. Here we expand around their initial values, but we could
  also have taken their average values over a long-term numerical
  integration (see, e.g.,~\cite{SanLocGio-2011a}).} and perform a
translation, $\Tscr_{F}$, defined as
$$
\Lvec=\Lambdavec-\Lambdavec^*\ .  
$$
This canonical transformation leaves the coordinates $\lambdavec$,
$\xivec$ and $\etavec$ unchanged.  The transformed Hamiltonian
$\Hscr^{(\Tscr)} = \Tscr_{F}(F)$ can be expanded in power series of
$\Lvec$, $\xivec$ and $\etavec$ around the origin.  Forgetting an
unessential constant, we rearrange the Hamiltonian of the system as
\begin{equation}
\Hscr^{(\Tscr)}=\nvec^*\cdot\Lvec+\sum_{j_1=2}^{\infty} h_{j_1,0}^{{\rm (Kep)}}(\Lvec)+
\mu\sum_{j_1=0}^{\infty}\sum_{j_2=0}^{\infty} h^{(\Tscr)}_{j_1,j_2}(\Lvec,\lambdavec,\xivec,\etavec)\ ,
\label{eq:H_trasl}
\end{equation}
where the functions $h^{(\Tscr)}_{j_1,j_2}$ are homogeneous
polynomials of degree $j_1$ in the fast actions $\Lvec$, degree $j_2$
in the secular variables $(\xivec,\etavec)$, and depend analytically
and periodically on the angles $\lambdavec$. The terms
$h_{j_1,0}^{{\rm (Kep)}}$ of the keplerian part are homogeneous
polynomials of degree $j_1$ in the fast actions $\Lvec$. We also
expand $h^{(\Tscr)}_{j_1,j_2}$ in Fourier series of the angles
$\lambdavec$.  In the latter equation, the term which is both linear
in the actions and independent of all the other canonical variables
(i.e.,~$\nvec^*\cdot \Lvec$) has been isolated in view of its
relevance in perturbation theory, as it will be discussed in the next
section.

All the expansions were carried out using a specially devised
algebraic manipulator (see \cite{GioSan-2012}).  In our computations
we truncate the expansion as follows. The keplerian part is expanded
up to the quadratic terms. The terms $h^{(\Tscr)}_{j_1,j_2}$ include
the linear terms in the fast actions $\Lvec$, all terms up to degree
$12$ in the secular variables $(\xivec,\etavec)$ and all terms up to
the trigonometric degree $12$ with respect to the angles $\lambdavec$.
The choice of the limits in the expansion is uniform for all the
systems that will be considered.  However, as we will explain in the
next section, the actual limits for the computation of the secular
approximation will be chosen as the lowest possible orders in
$\lambdavec$ and $(\xivec,\etavec)$, so as to include the main effects
of the proximity to a mean-motion resonance.

\section{Secular Hamiltonian}\label{sec:secular}
In this section we discuss the procedure for computing the secular
Hamiltonian via elimination of the fast angles. The classical
approach, usually found in the literature, consists in replacing the
Hamiltonian $\Hscr^{(\Tscr)}$, defined in~\eqref{eq:H_trasl}, by
\begin{equation}
\overline\Hscr = \frac{1}{4\pi^2}
\int_{0}^{2\pi}\!\!\!\int_{0}^{2\pi} \Hscr^{(\Tscr)}
\,{\rm d}\lambda_1\,{\rm d}\lambda_2\ .
\label{eq:average}
\end{equation}
The idea is that the effects due to the fast angles are negligible on
the long-term evolution and this averaged Hamiltonian represents a
``good approximation'' of the secular dynamics. This approach has been
critically considered by Arnold, quoting his book (i.e.,
\cite{Arnold-1989}, Chapter 10) {\it ``this principle is neither a
  theorem, an axiom, nor a definition, but rather a physical
  proposition, i.e., a vaguely formulated and, strictly speaking,
  untrue assertion. Such assertion are often fruitful sources of
  mathematical theorems.''}.

The secular Hamiltonian obtained in this way is the so-called
approximation at {\it order one in the masses} (or ``averaging by
scissors'') and is the basis of the Laplace-Lagrange theory for the
secular motion of perihelia and nodes of the planetary orbits. This
approximation corresponds to fixing the value of $\Lambdavec$, that
remains constant under the flow, and thereby the semi-major axes. The
averaged Hamiltonian, depending only on the secular variables, reduces
the problem to a system with two degrees of freedom.

Let us remark that the Laplace-Lagrange secular theory was developed
just considering the linear approximation of the dynamical
equations. An extension of the Laplace-Lagrange theory for
extrasolar systems, including also terms of higher order in the
eccentricities, can be found in~\cite{LibHen-2005,LibHen-2006}, where
the authors show that a secular Hamiltonian at order one in the masses
gives an accurate description of the long-term behavior for systems
which are not close to a mean-motion resonance.

One of the main achievements of the Laplace-Lagrange secular theory is
the explanation of the ``great inequality'' between Jupiter and
Saturn.  Indeed, they have shown that the near commensurability of the
two mean-motions (the $5$:$2$ near resonance) has a great impact on
the long-term behavior of the Solar system.  For that reason, a good
description of the secular dynamics of an exoplanetary system should
include a careful treatment of the influence of mean-motion resonances
on the long-term evolution.

Our purpose is to consider a secular Hamiltonian at {\it order two in
  the masses}.  The idea is to remove perturbatively the dependency on
the fast angles from the Hamiltonian~\eqref{eq:H_trasl}, considering
terms up to order two in the masses. This can be done using the
classical generating functions of the Hamilton-Jacobi formalism. Here
instead we use the Lie series formalism and implement the procedure in
a way that takes into account the Kolmogorov algorithm for the
construction of an invariant torus (see, \cite{Kolmogorov-1954}).
This is only a technical point and does not affect the results, our
choice is a question of convenience since the Lie series approach is a
direct method and is much more effective from the computational point of
view (see, e.g.,~\cite{GioLoc-2003}).

\subsection{Approximation at order two in the masses}\label{sbs:ord2}
Let us recall that in the expansion of the Hamiltonian
$\Hscr^{(\Tscr)}$, see~\eqref{eq:H_trasl}, the perturbation is of
order one in the masses and it is a polynomial in $\Lvec$, $\xivec$
and $\etavec$, and a trigonometric polynomial in the fast angles
$\lambdavec$.  We remove part of the dependence on the fast angles
performing a ``Kolmogorov-like'' normalization step, in the following
sense.  The suggestion of Kolmogorov is to give the Hamiltonian the
normal form $H(\Lvec,\lambdavec)= \nvec^*\cdot\Lvec + \Oscr(\Lvec^2)$,
for which the existence of an invariant torus $\Lvec=0$ is evident
(where we consider the secular variables just as parameters).  We give
the Hamiltonian the latter form up to terms of order $\Oscr(\mu^2)$.
More precisely, we perform a canonical transformation which removes
the dependence on the fast angles from terms which are independent of
and linear in the fast actions $\Lvec$ (i.e.,
equations~\eqref{eq:chi_1} and~\eqref{eq:chi_2}, respectively).
Therefore we replace the circular orbits of the Laplace-Lagrange
theory with an approximate invariant torus, thus establishing a better
approximation as the starting point of the classical theory.  We also
take into account the effects of near mean-motion resonances by
including in the averaging process the corresponding resonant
harmonics, as will be detailed hereafter.  The procedure is a little
cumbersome, and here we give only a sketch of the main path.  For a
detailed exposition one can refer to~\cite{LocGio-2007}
and~\cite{SanLocGio-2011a}.

The expansion of the Hamiltonian $\Hscr^{(\Tscr)}$,
see~\eqref{eq:H_trasl}, in view of the d'Alembert rules (see,
e.g.,~\cite{Poincare-1905}; see also
\cite{Kholshevnikov-1997,Kholshevnikov-2001} for a modern approach),
contains only specific combinations of terms.  Let us consider the
harmonic $\kvec \cdot \lambdavec$, where $\kvec=(k_1,\,k_2)$, and
introduce the so-called ``characteristic of the inequality''
$$
\Cscr_\Iscr(\kvec) = k_1 + k_2\ .
$$
The degree in the secular variables of the non-zero terms
appearing in the expansion must have the same parity of
$\Cscr_\Iscr(\kvec)$ and is at least equal to
$|\Cscr_\Iscr(\kvec)|\,$.

It is well known that the terms of the expansion that have the main
influence on the secular evolution are the ones related to low order
mean-motion resonances.  Therefore, if the ratio ${n_2^*}/{n_1^*}$ is
close to the rational approximation ${k_1^*}/{k_2^*}$, then the
effects due to the harmonics $(k_1^* \lambda_1 -k_2^* \lambda_2)$
should be taken into account in the secular Hamiltonian.  Let us also
recall that the coefficients of the Fourier expansion decay
exponentially with $|\kvec|_1=|k_1|+|k_2|$, so we just need to
consider low order resonances.

Let us go into details. Consider a system close to the $k_2^*:k_1^*$
mean-motion resonance and define the vector $\kvec^* = (k_1^*,
-k_2^*)$ and two integer parameters $K_F=|\kvec^*|_1$ and $K_S =
|\Cscr_\Iscr(\kvec^*)|$.  We denote by $\lceil f
\rceil_{\lambdavec;K_F}$ the Fourier expansion of a function $f$
truncated in such a way that we keep only the harmonics satisfying the
restriction $0<|\kvec|_1\leq K_F$. The effect of the near mean-motion
resonances is taken into account by choosing the parameters $K_S$ and
$K_F$ as the lowest limits that include the corresponding resonant
harmonics.  Using the Lie series algorithm to calculate the canonical
transformations (see, e.g.,~\cite{Henrard-1973} and
\cite{Giorgilli-1995}), we transform the
Hamiltonian~\eqref{eq:H_trasl} as $\widehat\Hscr^{(\Oscr 2)}=\exp
\lie{\mu\,\scriptchi_{1}^{(\Oscr 2)}}\,\Hscr^{(\Tscr)}$, with the
generating function $\mu\, \dispchi_1^{(\Oscr
  2)}(\lambdavec,\xivec,\etavec)$ determined by solving the equation
\begin{equation}
\sum_{j=1}^{2}n^*_j \frac{\partial\,\dispchi_{1}^{(\mathcal{O} 2)}}{\partial \lambda_j}
+\sum_{j_2=0}^{K_S}\left\lceil h_{0,j_2}^{(\Tscr)} \right\rceil_{\lambdavec;K_F}
(\lambdavec,\xivec,\etavec)=0\ .
\label{eq:chi_1}
\end{equation}
Notice that, by definition, the average over the fast angles of
$\left\lceil h_{0,j_2}^{(\Tscr)}\right\rceil_{\lambdavec;K_F}$ is
zero, which assures that~\eqref{eq:chi_1} can be solved provided that
the frequencies are non resonant up to order $K_F$.  The Hamiltonian
$\widehat {\mathcal H}^{(\mathcal{O} 2)}$ has the same form as
$\Hscr^{(\Tscr)}$ in~\eqref{eq:H_trasl}, with the functions
$h^{(\Tscr)}_{j_1,j_2}$ replaced by new ones, $\hat h^{(\mathcal{O}
  2)}_{j_1,j_2}$, generated by expanding the Lie series $\exp
\lie{\mu\,\scriptchi_{1}^{(\mathcal{O} 2)}}\, \mathcal{H}^{(\Tscr)}$
and gathering all the terms having the same degree both in the fast
actions and in the secular variables.

We now perform a second canonical transformation $\Hscr^{(\Oscr
  2)}=\exp \lie{\mu\,\scriptchi_{2}^{(\Oscr
    2)}}\,\widehat\Hscr^{(\Oscr 2)}$, where the generating function
$\mu\,\dispchi_{2}^{(\Oscr 2)}(\Lvec,\lambdavec,\xivec,\etavec)$,
which is linear with respect to $\Lvec$, is determined by solving the
equation
\begin{equation}
  \sum_{j=1}^{2}n^*_j \frac{\partial\,\dispchi_{2}^{(\Oscr
      2)}}{\partial \lambda_j} +\sum_{j_2=0}^{K_S}\left\lceil
  \hat{h}_{1,j_2}^{(\Oscr 2)} \right\rceil_{\lambdavec;K_F}
  (\Lvec,\lambdavec,\xivec,\etavec)=0\ .
  \label{eq:chi_2}
\end{equation}
Again,~\eqref{eq:chi_2} can be solved provided the frequencies are non
resonant up to order $K_F$ and the Hamiltonian
$\mathcal{H}^{(\mathcal{O} 2)}$ can be written in a form similar
to~\eqref{eq:H_trasl}, namely
\begin{equation}
  \Hscr^{(\Oscr 2)}(\Lvec,\lambdavec,\xivec,\etavec)=
  \nvec^*\cdot\Lvec+
  \sum_{j_1=2}^\infty h_{j_1,0}^{({\rm Kep})}(\Lvec)+
  \mu\sum_{j_1=0}^\infty\sum_{j_2=0}^\infty
  h_{j_1,j_2}^{(\Oscr 2)}(\Lvec,\lambdavec,\xivec,\etavec;\mu)+\Oscr(\mu^3)\ ,
  \label{eq:H_ord2}
\end{equation}
where the new functions $h^{(\Oscr 2)}_{j_1,j_2}$ are computed as
previously explained for $\hat h^{(\Oscr 2)}_{j_1,j_2}$ and still have
the same dependence on their arguments as $h^{(\Tscr)}_{j_1,j_2}$
in~\eqref{eq:H_trasl}. As we are interested in a second order
approximation, we have neglected the contribution of the order
$\Oscr(\mu^3)$ in the canonical transformations associated to
\eqref{eq:chi_1} and \eqref{eq:chi_2}. Following a common practice in
perturbation theory, we denote again by
$(\Lvec,\lambdavec,\xivec,\etavec)$ the transformed coordinates.

In the following, we will denote by $\Tscr_{\Oscr 2}$
the canonical transformation induced by the generating functions
$\mu\,\dispchi_{1}^{(\Oscr 2)}$ and $\mu\,\dispchi_{2}^{(\Oscr 2)}$,
namely
\begin{equation}
\Tscr_{\Oscr 2}(\Lvec,\lambdavec,\xivec,\etavec) = 
\exp \lie{\mu\,\scriptchi_{2}^{(\Oscr 2)}} \circ
\exp \lie{\mu\,\scriptchi_{1}^{(\Oscr 2)}}(\Lvec,\lambdavec,\xivec,\etavec)\ .
\label{eq:trasf_ord2}
\end{equation}

Let us remark that the non resonant condition
$$
\kvec\cdot\nvec^* \neq0\ ,
\qquad\hbox{for}\ \ 0<|\kvec|_1\leq K_F\ ,
$$
does not imply that the canonical change of coordinates is
convergent.  Indeed, the terms $\kvec\cdot\nvec^*$ that appear as the
denominators of the generating functions, even if they do not vanish,
can produce the so-called {\it small divisors}.  It is well known that
the presence of small divisors is a major problem in perturbation
theory.  Therefore, for each system considered in this work, we
check that the canonical transformation $\Tscr_{\Oscr 2}$ is near to the
identity and only in that case we proceed computing the approximation
at order two in the masses.

\subsection{Averaged Hamiltonian in diagonal form}\label{sbs:diag}
Starting from the Hamiltonian $\Hscr^{(\Oscr 2)}$ in~\eqref{eq:H_ord2},
we just need to perform an average over the fast angles $\lambdavec$.
More precisely, we consider the averaged Hamiltonian
\begin{equation}
\Hscr^{({\rm sec})}(\xivec,\etavec) = \left\langle \Hscr^{(\Oscr
  2)}\big|_{\Lvec=0}\right\rangle_{\lambdavec}\ ,
\label{eq:H_sec}
\end{equation}
namely we set $\Lvec = 0$ and average $\Hscr^{(\Oscr 2)}$ by removing
all the Fourier harmonics depending on the angles.  This results in
replacing the orbit having zero eccentricity with an invariant torus
of the unperturbed Hamiltonian.  The Hamiltonian so constructed is the
secular one, describing the slow motion of the eccentricities and
pericenters.  Concerning the approximation at order one in the masses,
let us recall that we directly average the Hamiltonian
$\Hscr^{(\Tscr)}$, see equation~\eqref{eq:average}.

After the averaging over the fast angles, the secular Hamiltonian has
two degrees of freedom and, in view of the d'Alembert rules, contains
only terms of even degree in $(\xivec,\etavec)$.  Therefore, the
lowest order approximation of the secular Hamiltonian, namely its
quadratic part, is essentially the one considered in the
Laplace-Lagrange theory.  The origin $(\xivec, \etavec) = (0, 0)$ is
an elliptic equilibrium point, and it is well known that one can find
a linear canonical transformation
$(\xivec,\etavec)=\Dscr(\xvec,\yvec)$ which diagonalizes the quadratic
part of the Hamiltonian, so that we may write $\Hscr^{({\rm sec})}$ in
the new coordinates as
\begin{equation}
H^{({\rm sec})}(\xvec, \yvec)=
\sum_{j=1}^{2}\nu_j\frac{x_j^2+y_j^2}{2}+
H_{2}^{(0)}(\xvec,\yvec)+H_{4}^{(0)}(\xvec,\yvec)+
\ldots\ ,
\label{eq:H-secdiag}
\end{equation}
where $\nu_j$ are the secular frequencies in the small oscillations
limit and $H_{2s}^{(0)}$ is a homogeneous polynomial of degree $2s+2$
in $(\xvec,\yvec)\,$.

To illustrate the transformations described hereabove, the secular
Hamiltonian, $\Hscr^{({\rm sec})}$, of the $\upsilon$ Andromedae
extrasolar system (see Section \ref{sec:andromedae} for a detailed
description of the system and a discussion on its proximity to the
$5$:$1$ mean-motion resonance) is reported in
appendix~\ref{app:sec}. First and second order approximations in the
masses, including terms up to order $6$ in $(\xivec,\etavec)$, are
given.

\section{Secular evolution in action-angle coordinates}\label{sec:actang}
Following \cite{LibHen-2006}, we now aim to introduce an action-angle
formulation, since its associated equations of motion are extremely
simple and can be integrated analytically. The secular
Hamiltonian~\eqref{eq:H-secdiag} has the form of a perturbed system of
harmonic oscillators, and thus we can construct a Birkhoff normal form
(see \cite{Birkhoff-1927}) introducing action-angle coordinates for
the secular variables, by means of Lie series (see,
e.g.,~\cite{Hori-1966,Deprit-1969,Giorgilli-1995}). Finally, an
analytical integration of the action-angle equations will allow us to
check the accuracy of our secular approximation, by comparing it to a
direct numerical integration of the Newton equations.

\subsection{Birkhoff normal form via Lie series}\label{sbs:birk}
As the construction of the Birkhoff normal form via Lie series is
explained in detail in many previous studies, here we just briefly
recall it, adapted to the present context.

First, we define a canonical transformation
$(\xvec,\yvec)=\Ascr(\Ivec, \phivec)$ introducing the usual
action-angle variables
\begin{equation}
x_j = \sqrt{2 I_j} \cos{\phi_j}\ ,\qquad
y_j = \sqrt{2 I_j} \sin{\phi_j}\ ,\qquad
j=1,\,2\ .
\end{equation}
The secular Hamiltonian in these variables reads
\begin{equation}
H^{({\rm sec})}(\Ivec, \phivec)= \nuvec\cdot\Ivec+
H_{2}^{(0)}(\Ivec, \phivec)+H_{4}^{(0)}(\Ivec, \phivec)+\ldots\ .
\label{eq:H-secdiag1}
\end{equation}

In order to remove the dependency of the secular angles $\phivec$ in
this expression, we compute the Birkhoff normal form up to order $r$,
\begin{equation}
H^{\,(r)} = Z_{0}(\Ivec)+\ldots+Z_{r}(\Ivec) + \Rscr^{(r)}(\Ivec,
\phivec)\ ,
\label{eq:H_r}
\end{equation}
where $Z_{s}$, for $s=0,\,\ldots\,,\,r\,$, is a homogeneous polynomial
of degree $s/2+1$ in $\Ivec$ and is zero for odd $s$. Only the
remainder, $\Rscr^{(r)}(\Ivec,\phivec)$, depends also on the angles
$\phivec$. Again, with a little abuse of notation, we denote by
$(\Ivec,\phivec)$ the new coordinates.  At each order $s>0$, we
determine the generating function $X^{(s)}$, by solving the equation
\begin{equation}
\left\{X^{(s)}\,,\,\nuvec\cdot\Ivec \right\}+
H_{s}(\Ivec, \phivec) = Z_{s}(\Ivec)\ .
\label{eqperchir+1}
\end{equation}

\noindent
Using the Lie series, we calculate the new Hamiltonian as
$H^{(s+1)}=\exp\lie{X^{(s+1)}}\,H^{(s)}$, provided that the
non-resonance condition
\begin{equation}
\kvec\cdot\nuvec\neq 0
\qquad\hbox{for}\ 
\kvec\in\Zbb^2\ {\rm such\ that}\ 0<|\kvec|_1\le s+2
\label{eq:nnres}
\end{equation}
is fulfilled.

Let us remark that, considering the Hamiltonian at order two in the
masses, the Birkhoff normal form is not always convergent at high
order, especially when the eccentricities are significant or the
system is {\it too close} to a mean-motion resonance.  Indeed, in
these cases, the transformation $\Tscr_{\Oscr 2}$, which brings the
Hamiltonian at order two in the masses, induces a big change in the
coefficients of the secular Hamiltonian, that can prevent the
convergence of the normalization procedure. On the contrary, the
algorithm seems to be convergent at first order in the masses (see the
{\it convergence au sens des astronomes} in \cite{LibHen-2006}).

Assuming that the non-resonance conditions~\eqref{eq:nnres} are
satisfied up to an order $r$ large enough, the remainder
$\Rscr^{(r)}(\Ivec, \phivec)$ can be neglected and we easily obtain an
analytical expression of the secular frequencies.  Indeed, the
equations of motion for the truncated Hamiltonian are
\begin{equation}
\dot\Ivec=0 \qquad {\rm and} \qquad \dot\phivec=
\frac{\partial H^{(r)}}{\partial\,\Ivec}\ ,
\label{eq:motion}
\end{equation}
and lead immediately to the expression of the two frequencies
$\dot\phi_1$ and $\dot\phi_2\,$.  Let us remark that, as the generating
functions of the Lie series depend only on the angular difference
$\phi_1 - \phi_2\,$, the frequency of the apsidal difference $\Delta
\varpi = \omega_1-\omega_2$ is equal to $\dot\phi_1 - \dot\phi_2\,$.

\subsection{Analytical integration}\label{sbs:an_int}
Using the equations in~\eqref{eq:motion}, we can compute the long-term
evolution on the secular invariant torus, namely
$$
\Ivec(t) = \Ivec(0)\qquad\hbox{and}\qquad
\phivec(t)=\phivec(0)+t\,\dot\phivec(0)\ ,
$$
where $\Ivec(0)$ and $\phivec(0)$ correspond to the values of the
initial conditions.  To validate our results, we will compare our
analytical integration with the direct numerical integration of the
full three-body problem, by using the symplectic integrator
$\Sscr\Bscr\Ascr\Bscr3$ (see \cite{LasRob-2001}).

Here we briefly explain how the analytical computation of the secular
evolution of the orbital elements is performed. Let us denote by
$\Tscr_{\Bscr}^{(r)}$ the canonical transformation induced by the
Birkhoff normalization up to the order $r$, namely
\begin{equation}
  \Tscr_{\Bscr}^{(r)}\big(\Ivec, \phivec\big) =
  \exp\lie{X^{(r)}}\circ\,\ldots\circ\,\exp\lie{X^{(1)}}
  \>\big(\Ivec, \phivec\big)\ .
\label{eq:birk_trasf}
\end{equation}
We denote by $\Cscr^{(r)}$ the composition of all the canonical
changes of coordinates defined in
Sections~\ref{sec:expansion}--\ref{sec:actang}, namely
\begin{equation}
\Cscr^{(r)}=\Tscr_{F}\circ\Tscr_{\Oscr 2}\circ\Dscr\circ\Ascr\circ\Tscr_{\Bscr}^{(r)}\ .
\label{def-Cscr}
\end{equation}
Taking the initial conditions
$\big(\avec(0),\lambdavec(0),\evec(0),\omegavec(0)\big)$, we can
compute the evolution of the orbital elements by using the following
scheme
\begin{equation}
\vcenter{\openup1\jot\halign{
 \hbox to 25 ex{\hfil $\displaystyle {#}$\hfil}
&\hbox to 20 ex{\hfil $\displaystyle {#}$\hfil}
&\hbox to 35 ex{\hfil $\displaystyle {#}$\hfil}\cr
\big(\avec(0),\lambdavec(0),\evec(0),\omegavec(0)\big)
&\build{\xrightarrow{\hspace*{25pt}}}_{}^{{{\displaystyle
\left(\Cscr^{(r)}\right)^{-1}\circ\Escr^{-1}}
\atop \phantom{0}}}
&\left({{\displaystyle \Ivec(0)}
\,,\, {\displaystyle \phivec(0)}}\right)
\cr
& &\bigg\downarrow
\cr
\big(\avec(t),\lambdavec(t),\evec(t),\omegavec(t)\big)
&\build{\xleftarrow{\hspace*{25pt}}}_{}^{{{\displaystyle
\Escr\circ\Cscr^{(r)}} \atop \phantom{0}}}
&\left({{\displaystyle \Ivec(t)=\Ivec(0)}
\,,\, {\displaystyle \phivec(t)=\phivec(0)+t\,\dot\phivec(0)}}\right)
\cr
}}
\ ,
\label{semi-analytical_scheme}
\end{equation}
where $(\Lambdavec,\lambdavec,\xivec,\etavec) =
\Escr^{-1}(\avec(0),\lambdavec(0),\evec(0),\omegavec(0))$ is the
non-canonical change of coordinates~\eqref{eq:poincvar}.  Thus, the
analytical integration via normal form actually reduces to a
transformation of the initial conditions to secular action-angles
coordinates, the computation of the flow at time $t$ in these
coordinates, followed by a transformation back to the original orbital
elements. Let us stress that, considering only the secular evolution,
the scheme above commutes only if $r$ is equal to infinity.

In the following sections, we will compare, for several extrasolar
systems, the analytical secular evolution of the eccentricities and
apsidal difference with the results of a direct numerical integration.
This kind of comparison has been shown to be a very stressing test
(see, e.g.,~\cite{LocGio-2007} and \cite{SanLocGio-2011a}) for the
accuracy of the whole algorithm constructing the normal form.

\section{Application to the $\upsilon$ Andromedae system}\label{sec:andromedae}
\begin{figure}[t]
  \begin{center}
    \includegraphics[width=0.8\textwidth]{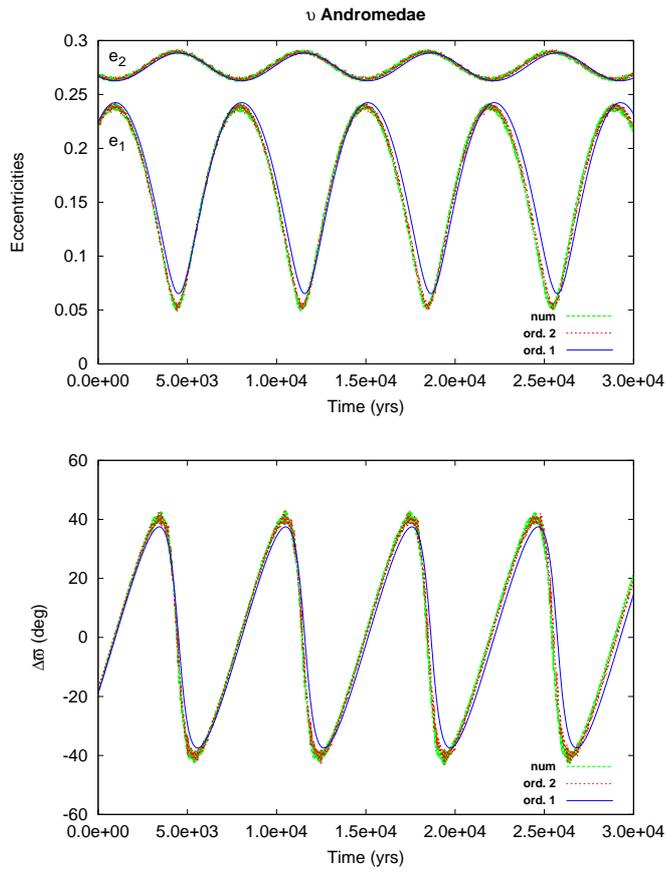}
  \end{center}
  \caption{Long-term evolution of the eccentricities (top panel) and
    difference of the longitudes of the pericenters (bottom panel) for
    the $\upsilon$ Andromedae system ($a_1/a_2=0.328$), obtained in
    three different ways: (i)~direct numerical integration via
    $\Sscr\Bscr\Ascr\Bscr3$ (green curves); (ii)~second order
    approximation (red curves); (iii)~first order approximation (blue
    curves). See text for more details.}
  \label{figandro}
\end{figure}

\begin{figure}[t]
  \begin{center}
    \includegraphics[width=\textwidth]{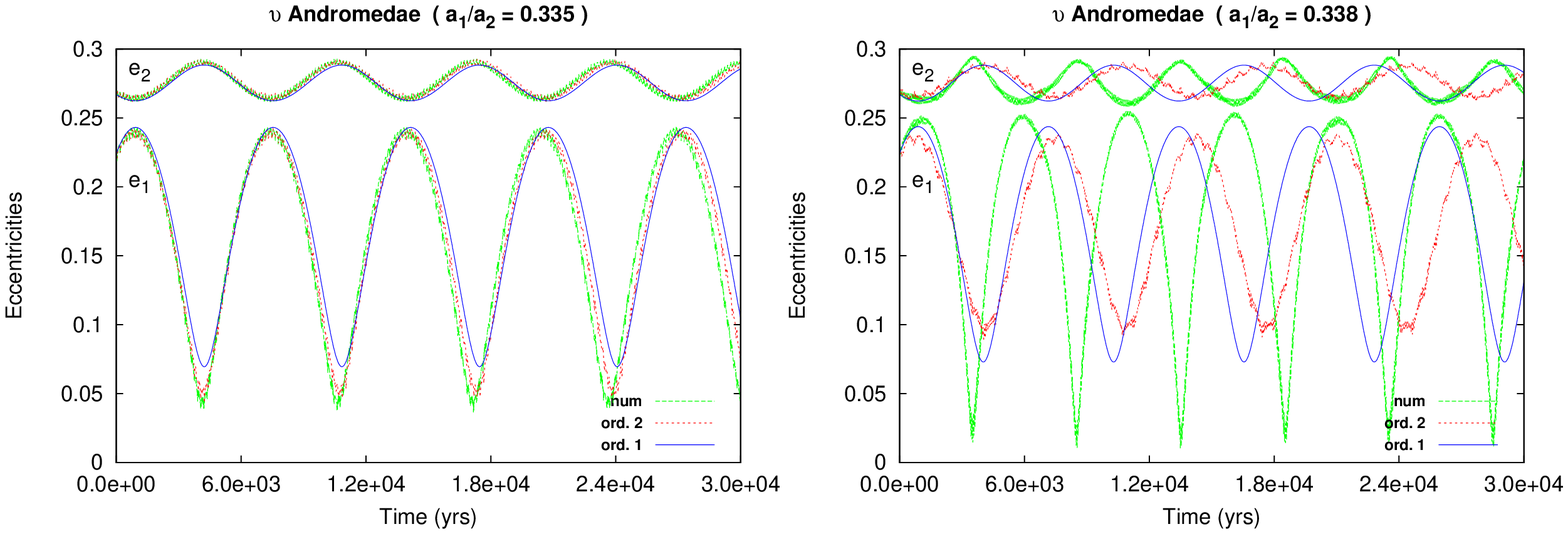}
  \end{center}
  \caption{Long-term evolution of slightly different versions of the
    $\upsilon$ Andromedae system, where the semi-major axis of planet {\it d}
    has been modified to be closer to the $5$:$1$ resonance. For
    $a_1/a_2=0.335$ (left panel), the secular approximation at order
    two in the masses is still efficient, while it is no more
    suitable for $a_1/a_2=0.338$ (right panel).}
 \label{figandro2}
\end{figure} 

We aim to investigate the improvements of the secular approximation at
order two in the masses, introduced in the previous sections, in
describing the long-term evolution of extrasolar systems close to a
mean-motion resonance. The planetary system $\upsilon$ Andromedae {\it
  c} and {\it d} is well known for his proximity to the $5$:$1$
mean-motion resonance. This has notably been confirmed analytically in
\cite{LibHen-2007}, where the authors argued that a first order
approximation gives a good {\it qualitative} approximation of the
secular dynamics of the system. In the following, we show that a
second order approximation can {\it quantitatively} enhance the
determination of the secular frequencies, as well as the extremal
values of the eccentricities and difference in apsidal angles reached
during the long-term evolution of the planets. For this study, we use
the orbital parameters reported in
\cite{WriUpaMarFis-2009}\footnote{Let us note that more recent
  parametrizations consistent with a $30^\circ$ mutual inclination of
  the two planets (\cite{McABenBarMarKorNelBut-2010}) and a fourth
  planet in the system (\cite{CurCanGeoChaPov-2010}) have been
  introduced.}.

In order to take into account the proximity of the system to the
$5$:$1$ mean-motion resonance, we must include, in the approximation
at order two in the masses, the effects of all the terms up to the
trigonometric degree $6$ in the fast angles and up to degree $4$ in
the secular variables, namely we set $K_F=6$ and $K_S=4$ (see the
definitions in Subsection 3.1).  After having constructed the secular
approximation, we perform a Birkhoff normal form up to order $r=10$
(see Subsection 4.1), which corresponds to taking into account the
secular variables up to order $12$.

In Figure~\ref{figandro}, we report the long-term evolution of the
eccentricities (top panel) and difference of the longitudes of
the pericenters (bottom panel) obtained analytically with our second
order approximation (red curves).  We compare it to the direct
numerical integration of the full three-body problem (i.e.,~including
the fast motions) in heliocentric canonical variables (green curves).
The agreement between both curves is excellent; the second order
theory reproduces qualitatively and quantitatively the results of the
numerical integration.  A comparison with the first order
approximation is also shown (blue curves) and gives evidence of the
improvement of the second order approximation for systems close to a
mean-motion resonance.

To highlight the dependency on the truncation parameters, we report,
in the table below, the values of the secular period for different
values of $K_F$ and $K_S$.  The period obtained via numerical
integration is $\sim 7000$ years.  As expected, higher values of the
truncation parameters allow to obtain better results, but with a
higher computational cost.  As already shown in Figure~\ref{figandro},
the main correction on the secular evolution is achieved when
considering the terms related to the $5$:$1$ mean-motion resonance,
i.e., $K_F=6$ and $K_S=4$. This validates our choice of the truncation
limits.
\begin{center}
\begin{tabular}{ccc}
\hline
$K_F$ & $K_S$ & Secular period  \\
\hline
$4$ & $2$ & $7132$ \\
$6$ & $4$ & $7035$ \\
$8$ & $6$ & $6998$ \\
\hline
\end{tabular}
\end{center}

In Figure~\ref{figandro2}, we slightly modify the semi-major axis of
planet {\it d} in such a way that the modified $\upsilon$ Andromedae
systems are closer and closer to the $5$:$1$ mean-motion
resonance. First we set $a_1/a_2=0.335$ (left panel).  In this case
the approximation at order one is not good enough, while the one at
order two is still suitable for the computation of the long-term
evolution of the system, even if the approximation is worst than the
one corresponding to the real $\upsilon$ Andromedae in
Figure~\ref{figandro}.  Finally, setting $a_1/a_2=0.338$ (right
panel), both the secular approximations completely fail. Indeed, in
this case, the system is {\it too close} to the resonance to be
qualitatively described by a secular approximation.

\section{Influence of the mean anomaly on the secular evolution}\label{sec:M}

\begin{figure}[t]
  \begin{center}
    \includegraphics[width=0.8\textwidth]{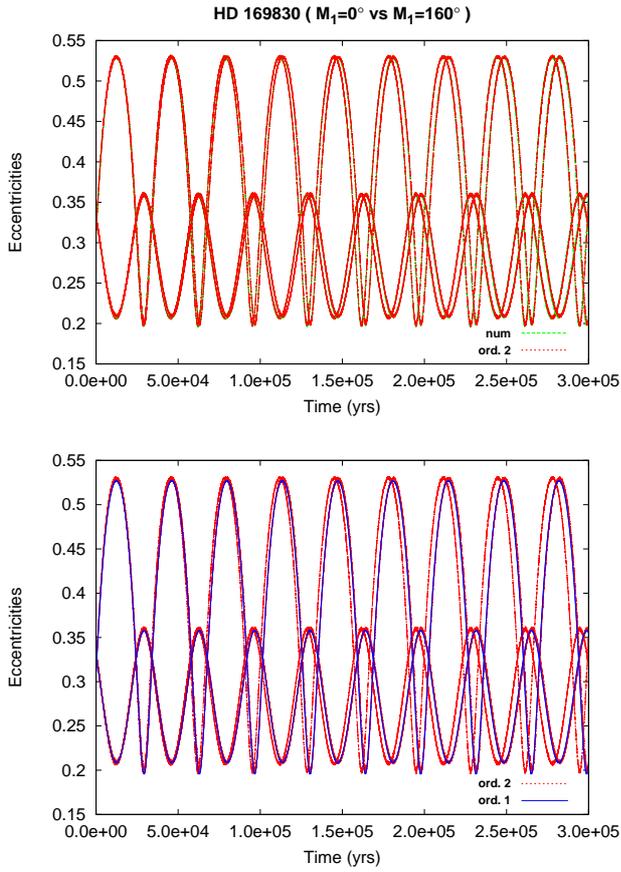}
  \end{center}
  \caption{Influence of the initial mean anomaly, $M$, on the secular
    evolution of the HD~169830 system. Top: long-term evolution for
    $M_1=0^\circ$ and $M_1=160^\circ$. In both cases, the second order
    approximation (red curves) reproduce accurately the numerical
    integration (green curves). Bottom: comparison between the
    evolution at order one (blue curve) and two (red curves) in the
    masses. See text for more details.}
 \label{figM}
 \end{figure} 

On the contrary to a first order analytical theory, an expansion to
the second order in the masses takes into account the influence of the
initial values of the fast angles on the secular evolution of the
system.  Let us stress that, as the averaging process giving the first
order secular Hamiltonian~\eqref{eq:average} does not involve any
canonical transformation, we take as ``averaged'' initial conditions
the original ones\footnote{For sake of completeness, we check that
  computing the ``averaged'' initial conditions using the generating
  functions $\chi_1$ and $\chi_2$, as in the approximation at order
  two in the masses, does not influence neither qualitatively nor
  quantitatively the results.}.

A change in the mean anomaly of a planet can have a significant impact
on the secular period of the system.  To show this, we plot, in
Figure~\ref{figM}, the extrasolar system HD~169830 for two different
values of the inner planet mean anomaly: $M_1=0^\circ$ and
$M_1=160^\circ$, all the other orbital parameters being unchanged and
issued from \cite{May-2004}. The displacement between the two secular
evolutions is obvious in the top panel. Let us note that, for both
values, our second order averaged Hamiltonian (in red) is very
accurate and coincides with the numerical evolution (in green). The
limitation of the secular expansion to order one in the masses is
pointed out in the bottom panel of Figure~\ref{figM}. The first order
evolution (blue curve) is the same regardless the initial value of the
mean anomaly, on the contrary to the approximation at order two in the
masses (red curves).

\section{Evaluation of the proximity to a mean-motion resonance}\label{sec:proxy}
We now study the proximity to a mean-motion resonance of the
two-planet exosystems discovered so far. This represents an extension
of the results in \cite{LibHen-2007}, previously obtained with an
approximation at order one in the masses.

Let us make some heuristic considerations. For systems that are {\it
  very close} to a low-order mean-motion resonance $k_2$:$k_1$,
i.e.,~$k_1 n_1^* -k_2 n_2^*\approx 0$, the generating functions
related to the second order approximation (i.e.,~$\mu\,
\dispchi_1^{(\Oscr 2)}$ and $\mu\, \dispchi_{2}^{(\Oscr 2)}$ defined
in~\eqref{eq:chi_1} and~\eqref{eq:chi_2}, respectively) contain the
so-called {\it small divisors}.  The presence of small divisors is a
major problem in perturbation theory, and here can prevent the
convergence of the second order averaging over the fast angles.
Instead, for a system that is only {\it near} to a mean-motion
resonance, but not too close, the approximation at order two in the
masses, including the main effects of the nearest low-order resonance,
enables to describe with great accuracy the long-term evolution of the
system. Finally, the secular evolution of a system that is {\it far}
from any low order mean-motion resonance is accurately depicted by the
approximation at order one in the masses.  Indeed, in this case, we
can safely replace the canonical transformation $\Tscr_{\Oscr 2}$ with
the classical first order average over the fast angles, see
equation~\eqref{eq:average}.

Let us go into details. To evaluate the proximity of a planetary
system to a mean-motion resonance, we introduce a criterion similar to
the one in \cite{LibHen-2007}.  The idea is to rate the proximity to a
mean-motion resonance by looking at the canonical change of
coordinates induced by the approximation at order two in the masses.
Precisely, we consider the low order terms of the canonical
transformation induced by $\Tscr_{\Oscr2}$, writing the averaged
variables $(\xivec',\etavec')$ as
\begin{align*}
\xi_j'&=\xi_j - \frac{\partial\,\mu\,\dispchi_1^{(\Oscr 2)}}{\partial\eta_j} =
\xi_j\,\left( 1-\frac{1}{\xi_j}\frac{\partial\,\mu\,\dispchi_1^{(\Oscr 2)}}{\partial\eta_j}\right)\ ,
\\
\eta_j'&=\eta_j - \frac{\partial\,\mu\,\dispchi_1^{(\Oscr 2)}}{\partial\xi_j} =
\eta_j\,\left( 1-\frac{1}{\eta_j}\frac{\partial\,\mu\,\dispchi_1^{(\Oscr 2)}}{\partial\xi_j}\right)\ ,
\end{align*}
for $j=1\,,\,2\,$.  The idea is that the generating function $\mu\,
\dispchi_1^{(\Oscr 2)}$ carries the main information about the
proximity to a mean-motion resonance, and we will focus here on the
coefficients of the functions
\begin{equation}
\delta\xi_j = \frac{1}{\xi_j}\frac{\partial\,\mu\,\dispchi_1^{(\Oscr 2)}}{\partial\eta_j}
\qquad\hbox{and}\qquad
\delta\eta_j = \frac{1}{\eta_j}\frac{\partial\,\mu\,\dispchi_1^{(\Oscr 2)}}{\partial\xi_j}\ .
\label{eq:prox}
\end{equation}

In these expressions, we aim to determine the most important periodic
terms whose corresponding harmonic $\kvec\cdot\lambdavec$ identifies
the main important mean-motion resonance to the system.  For each
system, we define a radius $\rhovec$ of a polydisk $\Delta_{\rhovec}$
around the origin of~$\Rbb^4$,
$$
\Delta_{\rhovec}=\left\{(\xivec,\etavec)\in\Rbb^4: \,\xi_j^2+\eta_j^2\leq
\rho_j^2\,,\ j=1,\,2 \right\}\ ,
$$
so as to include in that domain the initial conditions.  Given an
analytic function $f_{0,j_2}(\lambdavec,\xivec,\etavec)$ of the
form~\eqref{eq:H_trasl}, that reads
$$
f(\lambdavec,\xivec,\etavec) =\sum_\kvec
f^{(\kvec)}(\xivec, \etavec)\,{\sin\atop\cos}(\kvec\cdot\lambdavec)\ ,
$$
where
$$
f^{(\kvec)}(\xivec, \etavec)=
\sum_{|\lvec|+|\mvec|=j_2}
f^{(\kvec)}_{\lvec,\mvec}\,\xivec^{\lvec}\,\etavec^{\mvec}\ ,
$$
we can easily bound the sup-norm of the terms corresponding
to the harmonic $\kvec\cdot\lambdavec$ in $f$, by bounding
$f^{(\kvec)}$ in the polydisk $\Delta_{\rhovec}$ with the norm
\begin{equation}
\|f^{(\kvec)}\|_{\rhovec}=
\sum_{\lvec,\mvec}
|f^{(\kvec)}_{\lvec,\mvec}|\,\rho_1^{l_1+m_1}\,\rho_2^{l_2+m_2}\ .
\label{eq:norm}
\end{equation}
Applying the same criterion, for each angular combination
$\kvec\cdot\lambdavec$, we evaluate $\|\delta\xi^{(\kvec)}_j\|_{\rhovec}$
and $\|\delta\eta^{(\kvec)}_j\|_{\rhovec}$ for $j=1,\,2\,$, and, to
identify the closest mean-motion resonance to the system, we define
$$
\delta\xi_j^* = \max_{\kvec}(\|\delta\xi^{(\kvec)}_j\|_{\rhovec})
\qquad\hbox{and}\qquad \delta\eta_j^* =
\max_{\kvec}(\|\delta\eta^{(\kvec)}_j\|_{\rhovec})\ .
$$
For convenience we also introduce the following parameters: $\delta_j
= \min(\delta\xi_j^*, \delta\eta_j^*)$ for $j=1,2$ and
$\delta=\max(\delta_1, \delta_2)\,$. The parameter $\delta$ is a
measure of the change from the original secular variables to the
averaged ones.  The actual computation of $\delta$ is quite
cumbersome, but is more reliable than just looking at the semi-major
axes ratio, since it holds information about the non-linear
character of the system.

\renewcommand{\arraystretch}{1.4}
\begin{landscape}
\begin{table}
\vspace{5pt}
\caption{Evaluation of the proximity to a mean-motion resonance (MMR).
  We report here the values of $a_1/a_2$, $\mu$, $k_1n_1^*+k_2n_2^*$
  (where $k_1$ and $k_2$ correspond to the mean-motion resonance in
  brackets), $\delta_1$ and $\delta_2$ for each system considered.
  For our study, we use the following parametrizations:
  \cite{WriUpaMarFis-2009} for HD~11964, HD~12661, $\upsilon$
  Andromedae, HD~108874 and HD~183263; \cite{Mes-2011} for HD~74156
  and HD~177830; \cite{Jon-2010} for HD~134987; \cite{Gig-2011} for
  HD~163607; \cite{Seg-2009} for HD~147018; \cite{Tuo-2009} for
  HD~11506; \cite{Heb-2010} for HD~9446; \cite{May-2004} for
  HD~169830; JPL at the Julian Date~24404005 for the
  Sun-Jupiter-Saturn system; \cite{Wit-2009} for HD~128311.  See text
  for more details.}
\label{tab:prox}
\begin{center}
\begin{tabular}{cc|c|l|c|l|l|}
\cline{2-7}
\multicolumn{1}{c}{}&\multicolumn{1}{|c|}{System} & \hfil$a_1/a_2$& \hfil$\mu$ & \hfil$k_1n_1^*+k_2n_2^*$& \hfil$\delta_1$ & \hfil$\delta_2$  \\ \cline{1-7}
\multicolumn{1}{|c}{\multirow{6}{*}{\rotatebox{90}{Secular\ }}} &
\multicolumn{1}{|l|}{HD~11964} & $0.072$ & $5.4\times10^{-4}$ & $0.283\ (51$:$1)$ & 
$5.822\times10^{-4}\,\sin( -2\lambda_1  +\lambda_2)$
& $9.897\times10^{-4}\,\sin( -\lambda_1  +2\lambda_2)$
  \\ \cline{2-7}
\multicolumn{1}{|c}{}&\multicolumn{1}{|l|}{HD~74156} & $0.075$ & $6.3\times 10^{-3}$ & $0.579\ (48$:$1)$ &
$9.681\times10^{-4}\,\cos( \phantom{-}4\lambda_1  -\lambda_2)$
& $3.171\times10^{-4}\,\cos( \phantom{-}\lambda_1  -4\lambda_2)$
  \\ \cline{2-7}
\multicolumn{1}{|c}{}&\multicolumn{1}{|l|}{HD~134987} & $0.140$ &  $1.4 \times 10^{-3} $ & $0.052\ (19$:$1)$ &
$5.822\times10^{-4}\,\sin( -2\lambda_1  +\lambda_2)$
& $9.897\times10^{-4}\,\sin( -\lambda_1  +2\lambda_2)$
  \\ \cline{2-7}
\multicolumn{1}{|c}{}&\multicolumn{1}{|l|}{HD~163607} & $0.149$ & $2.0 \times 10^{-3}$ & $0.686\ (17$:$1)$ &
$1.376\times10^{-3}\,\cos( \phantom{-}3\lambda_1  -\lambda_2)$
& $3.492\times10^{-4}\,\sin( -\lambda_1  +2\lambda_2)$
  \\ \cline{2-7}
\multicolumn{1}{|c}{}&\multicolumn{1}{|l|}{HD~12661} & $0.287$ & $1.9 \times 10^{-3}$ & $0.671\ (6$:$1)$ &
$1.126\times10^{-3}\,\sin( -2\lambda_1  +\lambda_2)$
& $1.760\times10^{-3}\,\sin( -\lambda_1  +2\lambda_2)$
  \\ \cline{2-7}
\multicolumn{1}{|c}{}&\multicolumn{1}{|l|}{HD~147018} & $0.124$ & $6.8 \times 10^{-3}$ & $1.557\ (22$:$1)$ &
$2.455\times10^{-3}\,\sin( -2\lambda_1  +\lambda_2)$
& $1.658\times10^{-3}\,\sin( -\lambda_1  +2\lambda_2)$
  \\ \Cline{1.2pt}{1-7}
\multicolumn{1}{|c}{\multirow{6}{*}{\rotatebox{90}{near a MMR}}} &
\multicolumn{1}{|l|}{HD~11506} & $0.263$  &  $2.8 \times 10^{-3}$ &
$0.720\ (7$:$1)$ & 
$2.680\times10^{-3}\,\sin( -\lambda_1  +7\lambda_2)$
& $2.943\times10^{-3}\,\cos( \phantom{-}\lambda_1  -7\lambda_2)$
  \\ \cline{2-7}
\multicolumn{1}{|c}{}&\multicolumn{1}{|l|}{HD~177830} & $0.420$  & $9.7 \times 10^{-4}$ & $1.889\ (4$:$1)$ &
  $2.551\times10^{-3}\,\cos( \phantom{-}\lambda_1  -4\lambda_2)$
& $1.357\times10^{-3}\,\cos( \phantom{-}\lambda_1  -4\lambda_2)$
  \\ \cline{2-7}
\multicolumn{1}{|c}{}&\multicolumn{1}{|l|}{HD~9446} & $0.289$ & $1.7 \times 10^{-3}$ & $5.048\ (6$:$1)$ &
  $2.328\times10^{-3}\,\sin( -2\lambda_1  +\lambda_2)$
& $2.063\times10^{-3}\,\sin( -\lambda_1  +2\lambda_2)$
  \\ \cline{2-7}
\multicolumn{1}{|c}{}&\multicolumn{1}{|l|}{HD~169830} & $0.225$ & $2.8 \times 10^{-3}$ & $0.358\ (9$:$1)$ &
  $1.119\times10^{-2}\,\cos( \phantom{-}\lambda_1  -9\lambda_2)$
& $2.316\times10^{-2}\,\cos( \phantom{-}\lambda_1  -9\lambda_2)$
  \\ \cline{2-7}
\multicolumn{1}{|c}{}&\multicolumn{1}{|l|}{$\upsilon$ Andromedae } & $0.329$  & $3.0 \times 10^{-3}$ & $0.505\ (5$:$1)$ &
  $1.009\times10^{-2}\,\cos( \phantom{-}\lambda_1  -5\lambda_2)$
& $8.724\times10^{-3}\,\cos( \phantom{-}\lambda_1  -5\lambda_2)$
  \\ \cline{2-7}
\multicolumn{1}{|c}{}&\multicolumn{1}{|l|}{Sun-Jup-Sat} & $0.546$ & $9.5 \times 10^{-4}$  & $0.010\ (5$:$2)$ &
  $1.383\times10^{-2}\,\cos( -\lambda_1  +2\lambda_2)$
& $2.534\times10^{-2}\,\cos( 2\lambda_1  -5\lambda_2)$
  \\ \Cline{1.2pt}{1-7}
\multicolumn{1}{|c}{\multirow{3}{*}{\rotatebox{90}{MMR}}}&\multicolumn{1}{|l|}{HD~108874} & $0.380$ &  $1.3\times 10^{-3}$ & $0.338\ (4$:$1)$ &
  $1.052\times10^{-2}\,\cos( \phantom{-}\lambda_1  -4\lambda_2)$
& $4.314\times10^{-2}\,\sin( -\lambda_1 +4\lambda_2)$
  \\ \cline{2-7}
\multicolumn{1}{|c}{} &
\multicolumn{1}{|l|}{HD~128311} & $0.622$ & $3.7 \times 10^{-3}$ & $0.924\ (2$:$1)$&
$6.421\times10^{-1}\,\sin( -\lambda_1  +2\lambda_2)$
& $1.646\times10^{-1}\,\sin( -\lambda_1  +2\lambda_2)$
  \\ \cline{2-7}
\multicolumn{1}{|c}{}&\multicolumn{1}{|l|}{HD~183263} & $0.347$  & $3.1 \times 10^{-3}$ & $0.107\ (5$:$1)$ &
$2.772\times10^{-2}\,\cos( \phantom{-}\lambda_1  -5\lambda_2)$
& $5.253\times10^{-2}\,\cos( \phantom{-}\lambda_1  -5\lambda_2)$
  \\ \cline{1-7}
\end{tabular}
\end{center}
\end{table}
\end{landscape}

\begin{figure}[!h]
\begin{center}
\includegraphics[width=0.95\textwidth]{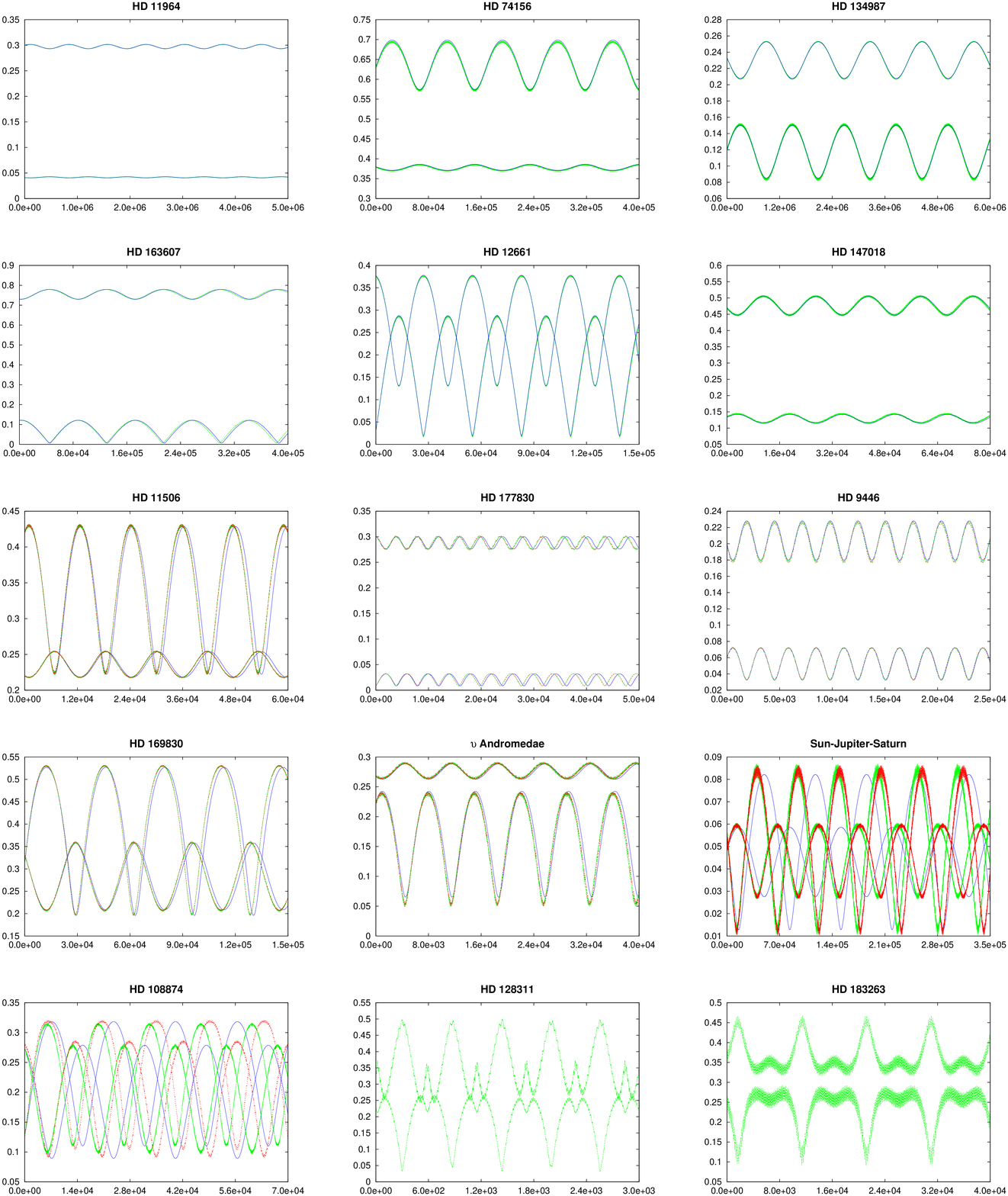}
\end{center}
\caption{Long-term evolution of the eccentricities of the extrasolar
  systems of Table \ref{tab:prox}, obtained by three different ways:
  (i)~the direct numerical integration via $\Sscr\Bscr\Ascr\Bscr3$
  (green curves); (ii)~the second order approximation (red curves);
  (iii)~the first order approximation (blue curves).  See text for
  more details.}
\label{fig:ecc}
\end{figure}

The results for all the extrasolar systems we have considered are
shown in Table~\ref{tab:prox} and Figure~\ref{fig:ecc}.  In
Table~\ref{tab:prox}, we report, for each system, the numerical values
of the semi-major axes ratio $a_1/a_2$, the mass ratio $\mu$, the
small divisor $k_1n_1^*+k_2n_2^*$ (where $k_1$ and $k_2$ correspond to
the mean-motion resonance in brackets) and the two aforementioned
parameters, $\delta_1$ and $\delta_2$, that will be used to set up a
criterion evaluating the proximity to a mean-motion resonance.  In
Figure~\ref{fig:ecc}, we plot the evolution of the eccentricities
obtained by direct numerical integration of the Newton equations
(green curves) and the ones obtained with a secular Hamiltonian at
order one (blue curves) and two (red curves) in the masses. As in
Section~\ref{sec:andromedae}, we limit the Birkhoff normal form at
order $r=10$ (i.e.,~$12$ in the secular variables). Let us stress
that, for the HD~128311 and HD~183236 systems, due to their close
proximity to the mean-motion resonances $2$:$1$ and $5$:$1$,
respectively, the canonical transformation $\Tscr_{\Oscr2}$ performing
the second order approximation is not close to the
identity. Therefore, we report only their numerical integrations.

Comparing the data in Table~\ref{tab:prox} and the corresponding plots
in Figure~\ref{fig:ecc}, we can roughly distinguish three cases: (i)
if $\delta<2.6\times10^{-3}$, the first order approximation describes
the secular evolution with great accuracy, therefore we label these
systems as {\it secular\/}; (ii) if
$2.6\times10^{-3}<\delta\leq2.6\times10^{-2}$, a second order average
of the Hamiltonian is required to describe the long-term evolution in
detail, we label them as {\it near mean-motion resonance} (the
$\upsilon$ Andromedae system analyzed in Section~\ref{sec:andromedae}
is the typical example of such category); (iii) if
$\delta>2.6\times10^{-2}$, the system is {\it too close} to a
mean-motion resonance and a secular approximation is not enough to
describe their evolution, then we label them as {\it in mean-motion
  resonance}. In this case it would be worthwhile to consider a
resonant Hamiltonian instead of a secular approximation.  Let us note
that the Sun-Jupiter-Saturn system and HD~108874 are both really close
to the border between {\it near mean-motion resonance} and {\it in
  mean-motion resonance} categories. Indeed, a much refined secular
approximation could be used, for instance increasing the values of
$K_F$ and $K_S$, without having to resort to a resonant model.

The criterion introduced above is clearly heuristic and quite rough,
nevertheless we think it is useful to discriminate between the
different behaviors of planetary systems.

\section{Conclusions and outlooks}\label{sec:results}
In this work we have analyzed the long-term evolution of several
exoplanetary systems by using a secular Hamiltonian at order two in
the masses.  The second order approximation, as explained in detail in
Section~\ref{sec:secular}, includes a careful treatment of the main
effects due to the proximity to a low-order mean-motion resonance.

Starting from the secular Hamiltonian, we have computed a high-order
Birkhoff normal form via Lie series, introducing action-angle
coordinates for the secular variables.  This enabled us to compute
analytically the evolution on the secular invariant torus and to
obtain the long-term evolution of the eccentricities and apsidal
difference.

As a result, for all the systems that are not too close to a
mean-motion resonance, we have shown an excellent agreement with the
direct numerical integration of the full three-body problem (including
the fast motions).  The influence of the mean anomalies on the secular
evolution of the systems has also been pointed out.  Furthermore,
evaluating the difference between the original and the averaged
secular coordinates, we have set up a simple (and rough) criterion to
discriminate between three different behaviors: (i)~secular system,
where a first order approximation is enough; (ii)~system near a
mean-motion resonance, where an approximation at order two is
required; (iii)~system that are really close to or in a mean-motion
resonance, where a resonant model should be used. In particular, we
find that HD~11964, HD~74156, HD~134987, HD~163607, HD~12661 and
HD~147018 belong to~(i); HD~11506, HD~177830, HD~9446, HD~169830 and
$\upsilon$~Andromedae to~(ii); HD~108874, HD~128311 and HD~183263
to~(iii).

Let us remark that these results could be extended to the spatial case
with minor changes.  Indeed, after the reduction of the angular
momentum, the starting Hamiltonian would have exactly the same form as
$\Hscr^{(\Tscr)}$, defined in~\eqref{eq:H_trasl}.

Moreover, having such a good analytical description of the orbits,
even for systems that are near a mean-motion resonance, we can also
study the effective stability of extrasolar planetary systems in the
framework of the KAM and Nekhoroshev theories. This topic deserves
further investigation in the future.

Finally, a natural extension to the present work would be the study of
the secular evolution of systems that are really close to or in a
mean-motion resonance.  As previously said, a resonant Hamiltonian
that keeps the dependency on the resonant combinations of the fast
angles has to be considered. This study is reserved for future work.

\begin{acknowledgements}
The work of A.-S.~L. is supported by an FNRS Postdoctoral Research
Fellowship. The work of M.~S. is supported by an FSR Incoming
Post-doctoral Fellowship of the Acad\'emie universitaire Louvain,
co-funded by the Marie Curie Actions of the European Commission.
\end{acknowledgements}

\appendix
\section{Secular Hamiltonian for the $\upsilon$~Andromedae up to order 6}\label{app:sec}
We report here the expansion of the secular Hamiltonian $\Hscr^{({\rm
    sec})}$ (equation~\eqref{eq:H_sec}) of the $\upsilon$ Andromedae
extrasolar system up to degree~$6$ in $(\xivec,\etavec)\,$. In
particular, we show both the approximations at order one and two in
the masses to highlight the differences. A detailed description of the
$\upsilon$ Andromedae system is given in Section~\ref{sec:andromedae}.
As this system is near the $5$:$1$ mean-motion resonance, the main
difference between the two secular approximations affects terms that
are at least of order $6$ in the canonical secular variables.
 
\renewcommand{\arraystretch}{1.2}
\begin{center}
  \begin{tabular}{|c|c|c|c|r|r|}
    \hline
    $\xi_1$ & $\xi_2$ & $\eta_1$ & $\eta_2$ & First order\hfil\hss & Second order\hfil\hss \\
    \hline
    0 &   0 &   0 &   0 & $-3.8449638957147059 \times 10^{+0}$ & $-3.8490132363346130 \times 10^{+0}$ \\
    \hline
    2 &   0 &   0 &   0 & $-4.7203675679835364 \times 10^{-4}$ & $-4.7442843563932181 \times 10^{-4}$ \\
    1 &   1 &   0 &   0 & $ 1.9765062410537654 \times 10^{-4}$ & $ 1.9478085580405423 \times 10^{-4}$ \\
    0 &   2 &   0 &   0 & $-1.2594397524843563 \times 10^{-4}$ & $-1.2389253809188814 \times 10^{-4}$ \\
    0 &   0 &   2 &   0 & $-4.7203675679835364 \times 10^{-4}$ & $-4.7442843563932181 \times 10^{-4}$ \\
    0 &   0 &   1 &   1 & $ 1.9765062410537654 \times 10^{-4}$ & $ 1.9478085580405423 \times 10^{-4}$ \\
    0 &   0 &   0 &   2 & $-1.2594397524843563 \times 10^{-4}$ & $-1.2389253809188814 \times 10^{-4}$ \\
    \hline
    4 &   0 &   0 &   0 & $ 1.4338305925091211 \times 10^{-4}$ & $ 1.4383176583648995 \times 10^{-4}$ \\
    3 &   1 &   0 &   0 & $ 4.3147125112054390 \times 10^{-4}$ & $ 4.5045949999300181 \times 10^{-4}$ \\
    2 &   2 &   0 &   0 & $-7.4883810227863515 \times 10^{-4}$ & $-7.5868060326294868 \times 10^{-4}$ \\
    2 &   0 &   2 &   0 & $ 2.8676611850182422 \times 10^{-4}$ & $ 2.8766295710810365 \times 10^{-4}$ \\
    2 &   0 &   1 &   1 & $ 4.3147125112054390 \times 10^{-4}$ & $ 4.5045950531661890 \times 10^{-4}$ \\
    2 &   0 &   0 &   2 & $-5.1509576520639426 \times 10^{-4}$ & $-5.2857952689768168 \times 10^{-4}$ \\
    1 &   3 &   0 &   0 & $ 3.3341302753346505 \times 10^{-4}$ & $ 3.2386648540035952 \times 10^{-4}$ \\
    1 &   1 &   2 &   0 & $ 4.3147125112054390 \times 10^{-4}$ & $ 4.5045950531661890 \times 10^{-4}$ \\
    1 &   1 &   1 &   1 & $-4.6748467414448156 \times 10^{-4}$ & $-4.6020211054162414 \times 10^{-4}$ \\
    1 &   1 &   0 &   2 & $ 3.3341302753346505 \times 10^{-4}$ & $ 3.2386626581402522 \times 10^{-4}$ \\
    0 &   4 &   0 &   0 & $-9.4514514989701095 \times 10^{-5}$ & $-9.0913589478614943 \times 10^{-5}$ \\
    0 &   2 &   2 &   0 & $-5.1509576520639426 \times 10^{-4}$ & $-5.2857952689768124 \times 10^{-4}$ \\
    0 &   2 &   1 &   1 & $ 3.3341302753346505 \times 10^{-4}$ & $ 3.2386626581402554 \times 10^{-4}$ \\
    0 &   2 &   0 &   2 & $-1.8902902997940219 \times 10^{-4}$ & $-1.8182716424233877 \times 10^{-4}$ \\
    \hline
  \end{tabular}
\end{center}

\begin{center}
  \begin{tabular}{|c|c|c|c|r|r|}
    \hline
    0 &   0 &   4 &   0 & $ 1.4338305925091211 \times 10^{-4}$ & $ 1.4383176583649006 \times 10^{-4}$ \\
    0 &   0 &   3 &   1 & $ 4.3147125112054390 \times 10^{-4}$ & $ 4.5045949999300165 \times 10^{-4}$ \\
    0 &   0 &   2 &   2 & $-7.4883810227863515 \times 10^{-4}$ & $-7.5868060326294889 \times 10^{-4}$ \\
    0 &   0 &   1 &   3 & $ 3.3341302753346505 \times 10^{-4}$ & $ 3.2386648540035947 \times 10^{-4}$ \\
    0 &   0 &   0 &   4 & $-9.4514514989701095 \times 10^{-5}$ & $-9.0913589478614848 \times 10^{-5}$ \\
    \hline
    6 &   0 &   0 &   0 & $ 8.0737006151169034 \times 10^{-5}$ & $ 1.3917499875750025 \times 10^{-4}$ \\
    5 &   1 &   0 &   0 & $-1.4728781329895123 \times 10^{-4}$ & $-5.7065127472031446 \times 10^{-4}$ \\
    4 &   2 &   0 &   0 & $-3.8625662439607426 \times 10^{-4}$ & $ 8.8051997114562091 \times 10^{-4}$ \\
    4 &   0 &   2 &   0 & $ 2.4221101845350710 \times 10^{-4}$ & $ 4.1753226095492534 \times 10^{-4}$ \\
    4 &   0 &   1 &   1 & $-1.4728781329895123 \times 10^{-4}$ & $-5.7066077231623241 \times 10^{-4}$ \\
    4 &   0 &   0 &   2 & $ 4.4154338663068642 \times 10^{-5}$ & $ 3.1718226339201008 \times 10^{-4}$ \\
    3 &   3 &   0 &   0 & $ 1.1715817984811095 \times 10^{-3}$ & $-9.4757874129889675 \times 10^{-4}$ \\
    3 &   1 &   2 &   0 & $-2.9457562659790246 \times 10^{-4}$ & $-1.1413685872968715 \times 10^{-3}$ \\
    3 &   1 &   1 &   1 & $-8.6082192611828580 \times 10^{-4}$ & $ 1.1266453489623077 \times 10^{-3}$ \\
    3 &   1 &   0 &   2 & $ 9.0270698998234857 \times 10^{-4}$ & $-7.3916625373992911 \times 10^{-4}$ \\
    2 &   4 &   0 &   0 & $-1.0274689835474350 \times 10^{-3}$ & $ 8.0672276683552634 \times 10^{-4}$ \\
    2 &   2 &   2 &   0 & $-3.4210228573300561 \times 10^{-4}$ & $ 1.1977229419632242 \times 10^{-3}$ \\
    2 &   2 &   1 &   1 & $ 1.7093314154786317 \times 10^{-3}$ & $-1.3644052823847199 \times 10^{-3}$ \\
    2 &   2 &   0 &   2 & $-1.4654674698375437 \times 10^{-3}$ & $ 1.2083393139899557 \times 10^{-3}$ \\
    2 &   0 &   4 &   0 & $ 2.4221101845350710 \times 10^{-4}$ & $ 4.1753226095492669 \times 10^{-4}$ \\
    2 &   0 &   3 &   1 & $-2.9457562659790246 \times 10^{-4}$ & $-1.1413685872968726 \times 10^{-3}$ \\
    2 &   0 &   2 &   2 & $-3.4210228573300561 \times 10^{-4}$ & $ 1.1977229419632145 \times 10^{-3}$ \\
    2 &   0 &   1 &   3 & $ 9.0270698998234857 \times 10^{-4}$ & $-7.3916701878536670 \times 10^{-4}$ \\
    2 &   0 &   0 &   4 & $-4.3799848629010854 \times 10^{-4}$ & $ 4.0161436876523369 \times 10^{-4}$ \\
    1 &   5 &   0 &   0 & $ 3.8723701961918303 \times 10^{-4}$ & $-2.8749294965146893 \times 10^{-4}$ \\
    1 &   3 &   2 &   0 & $ 9.0270698998234857 \times 10^{-4}$ & $-7.3916701878535239 \times 10^{-4}$ \\
    1 &   3 &   1 &   1 & $-1.1789409945146532 \times 10^{-3}$ & $ 8.1021818828667783 \times 10^{-4}$ \\
    1 &   3 &   0 &   2 & $ 7.7447403923836607 \times 10^{-4}$ & $-5.7498741718758071 \times 10^{-4}$ \\
    1 &   1 &   4 &   0 & $-1.4728781329895123 \times 10^{-4}$ & $-5.7066077231623220 \times 10^{-4}$ \\
    1 &   1 &   3 &   1 & $-8.6082192611828580 \times 10^{-4}$ & $ 1.1266453489623106 \times 10^{-3}$ \\
    1 &   1 &   2 &   2 & $ 1.7093314154786317 \times 10^{-3}$ & $-1.3644052823847082 \times 10^{-3}$ \\
    1 &   1 &   1 &   3 & $-1.1789409945146532 \times 10^{-3}$ & $ 8.1021818828667317 \times 10^{-4}$ \\
    1 &   1 &   0 &   4 & $ 3.8723701961918303 \times 10^{-4}$ & $-2.8749285273268994 \times 10^{-4}$ \\
    0 &   6 &   0 &   0 & $-6.9390702058934342 \times 10^{-5}$ & $ 6.9301447295937329 \times 10^{-5}$ \\
    0 &   4 &   2 &   0 & $-4.3799848629010854 \times 10^{-4}$ & $ 4.0161436876524556 \times 10^{-4}$ \\
    0 &   4 &   1 &   1 & $ 3.8723701961918303 \times 10^{-4}$ & $-2.8749285273267357 \times 10^{-4}$ \\
    0 &   4 &   0 &   2 & $-2.0817210617680304 \times 10^{-4}$ & $ 2.0789708715975659 \times 10^{-4}$ \\
    0 &   2 &   4 &   0 & $ 4.4154338663068642 \times 10^{-5}$ & $ 3.1718226339199647 \times 10^{-4}$ \\
    0 &   2 &   3 &   1 & $ 9.0270698998234857 \times 10^{-4}$ & $-7.3916625373990645 \times 10^{-4}$ \\
    0 &   2 &   2 &   2 & $-1.4654674698375437 \times 10^{-3}$ & $ 1.2083393139899626 \times 10^{-3}$ \\
    0 &   2 &   1 &   3 & $ 7.7447403923836607 \times 10^{-4}$ & $-5.7498741718756510 \times 10^{-4}$ \\
    0 &   2 &   0 &   4 & $-2.0817210617680304 \times 10^{-4}$ & $ 2.0789708715975702 \times 10^{-4}$ \\
    0 &   0 &   6 &   0 & $ 8.0737006151169034 \times 10^{-5}$ & $ 1.3917499875750112 \times 10^{-4}$ \\
    0 &   0 &   5 &   1 & $-1.4728781329895123 \times 10^{-4}$ & $-5.7065127472031587 \times 10^{-4}$ \\
    0 &   0 &   4 &   2 & $-3.8625662439607426 \times 10^{-4}$ & $ 8.8051997114559945 \times 10^{-4}$ \\
    0 &   0 &   3 &   3 & $ 1.1715817984811095 \times 10^{-3}$ & $-9.4757874129888428 \times 10^{-4}$ \\
    0 &   0 &   2 &   4 & $-1.0274689835474350 \times 10^{-3}$ & $ 8.0672276683552298 \times 10^{-4}$ \\
    0 &   0 &   1 &   5 & $ 3.8723701961918303 \times 10^{-4}$ & $-2.8749294965147088 \times 10^{-4}$ \\
    0 &   0 &   0 &   6 & $-6.9390702058934342 \times 10^{-5}$ & $ 6.9301447295938372 \times 10^{-5}$ \\
    \hline
  \end{tabular}
\end{center}

\end{document}